\documentclass[%
 reprint,
 amsmath,amssymb,
 aps,
]{revtex4-2}

\usepackage{graphicx}
\usepackage{dcolumn}
\usepackage{bm}
\usepackage{physics}
\usepackage{dsfont}

\newcommand{\LL}{\mathcal{L}}
\newcommand{\PP}{\mathcal{P}}
\newcommand{\QQ}{\mathcal{Q}}

\begin{document}

\title{Capturing non-Markovian polaron dressing
with the master equation formalism}
\author{Jake Iles-Smith}
\affiliation{Department of Physics and Astronomy, University of Manchester, Oxford Road, Manchester M13 9PL, United Kingdom}
\author{Owen Diba}
\affiliation{Department of Physics and Astronomy, University of Manchester, Oxford Road, Manchester M13 9PL, United Kingdom}
\author{Ahsan Nazir}
\affiliation{Department of Physics and Astronomy, University of Manchester, Oxford Road, Manchester M13 9PL, United Kingdom}


\begin{abstract}
Understanding the dynamics of open quantum systems in strong coupling and non-Markovian regimes
remains a formidable theoretical challenge.  
One popular and well-established method of approximation 
in these circumstances 
is provided by the polaron master equation (PME). 
In this work we reevaluate and extend the validity of the PME to capture the impact of non-Markovian polaron dressing, induced by non-equilibrium open system dynamics. 
By comparing with numerically exact techniques, 
we confirm that while the standard PME successfully predicts the 
dynamics of 
system observables that commute with the polaron transformation (e.g.~populations in the Pauli $z$-basis), it can struggle to fully capture those that do not (e.g.~coherences). 
This limitation stems from the 
mixing of system and environment degrees of freedom inherent to the polaron transformation, 
which affects the accuracy of calculated expectation values within the polaron frame. 
Employing the Nakajima-Zwanzig projection operator formalism, we introduce correction terms that provide an accurate description of 
observables that do not commute with the transformation. 
We demonstrate the significance of the correction terms in two cases, the canonical spin-boson model and a dissipative time-dependent Landau-Zener protocol, where they are shown to impact the system dynamics on both short and long timescales.

\end{abstract}

\maketitle

\section{Introduction}

The interaction between quantum systems and their surrounding 
environments is a fundamental aspect of quantum mechanics, 
leading to the dissipative and decoherent dynamics that are characteristic of 
most real physical 
situations. 
The complexity of these interactions, especially outside the 
perturbative regime of weak system-environment interactions, 
presents a significant theoretical challenge. 
In recent years, there has been substantial progress in 
overcoming these challenges through the development of 
efficient numerical algorithms, such as the hierarchical equations of 
motion~\cite{tanimura89heom,Ishizaki05heom}, the 
time evolved matrix product operator (TEMPO)~\cite{strathearn2018efficient,jorgensen2019exploiting,link2023open}, 
its generalisations~\cite{cygorek2022simulation}, 
and the time evolving density operator using orthogonal polynomials 
algorithm (TEDOPA)~\cite{prior2010efficient,chin2010exact,rosenbach2016efficient}.
While these methods have proven invaluable for studying open quantum systems in previously inaccessible 
regimes, 
they are not completely without limitation. For example, they 
are typically numerical black boxes, 
providing limited analytic insight into the evolution of the system 
and the environment.  
Furthermore, even with highly efficient tensor network representations, 
such techniques can still be computationally demanding, 
making accurate long time dynamics a challenge to obtain (as we shall see for a particular example below).

In contrast, 
the quantum master equation (QME) 
formalism offers an efficient and straightforward approach to simulating 
open quantum systems, with the potential to draw analytic insight from the resulting equations of motion.
While the Nakajima-Zwanzig projection operator procedure~\cite{nakajima58, Zwanzwig60} 
leads to a QME that is formally exact, for most models of interest 
it is necessary to make approximations to obtain a tractable theory. 
The 
standard procedure is to treat the system-environment interaction 
as a weak perturbation, leading to the well-known Redfield 
master equation~\cite{Redfield65,breuer2002theory}. 
However, there is now a 
significant body of work that extends QMEs beyond 
the weak coupling regime. 
Examples include the pseudo-mode~\cite{Garraway97pseudo,Garraway97decay,Dalton01pseudo} 
and reaction coordinate~\cite{iles2014environmental,nazir2018reaction} 
techniques, 
in which key environmental degrees of freedom are incorporated 
into 
the system of interest, which is then considered to be 
weakly coupled to the residual environment. 

In this work, we focus on another such QME treatment, 
known as the polaron master equation (PME)~\cite{nazir2016modelling}, 
in which a unitary transformation is used to dress the system 
with displacements of the environmental (vibrational) modes. 
This allows one to derive a Born-Markov master equation 
that is 
non-perturbative in the system-environment interaction strength, and thus valid over a larger regime of parameters than the usual Redfield master equation. 
The resulting theory and its variational extensions~\cite{Silbey84variational,mccutcheon11var} 
have been used extensively to study the behaviour of dissipative 
quantum systems, including semiconductor quantum 
dots~\cite{wilson02,mccutcheon2010quantum,iles2014environmental}, 
excitonic energy transfer in molecular 
complexes~\cite{jang08transfer,Nazir09transfer,jang2011theory,kolli2011electronic}, 
and quantum thermodynamics applications~\cite{gelbwaser2015strongly,krause2015thermodynamics,hsieh2019nonequilibrium,diba2024quantum}.

Here we show that although the PME 
can very accurately describe 
the evolution of system observables that commute with the 
polaron transformation (e.g. in our case populations in the Pauli $z$-basis), it fails to describe non-commuting observables, 
such as the coherence of the reduced state, to the same level of accuracy.
This is a consequence of the polaron transformation mixing the 
system and environment degrees of freedom, which must be accounted for 
when calculating expectation values in the lab frame, i.e.~the frame of reference in which our Hamiltonian is originally specified and thus assumed to be physically accessible. 
Using the Nakajima-Zwanzwig projection operator formalism 
we derive correction terms to non-commuting observables, 
which faithfully describe their behaviour when benchmarked against 
the numerically-exact TEMPO method.
Furthermore, the computational simplicity of the polaron approach 
allows the calculation of the long-time and steady-state behaviour of the system, which is extremely challenging with TEMPO due to the 
accumulation of numerical errors. 
We demonstrate the significance of the correction terms in two cases, 
the canonical spin-boson model and a dissipative time-dependent 
Landau-Zener (LZ) protocol.

The paper is organised as follows. In Sec.~\ref{sec:TCL} we give an outline of the Nakajima-Zwanzig projection operator technique~\cite{nakajima58, Zwanzwig60} and how it may be used to obtain the time-convolutionless (TCL) master equation. In Sec.~\ref{sec:polaron} we introduce the polaron transformation and derive a TCL master equation in the polaron frame. In Sec.~\ref{sec:lab_frame} we derive correction terms for calculating lab-frame expectation values with the polaron frame master equation. Sec.~\ref{sec:tempo} presents an outline of the TEMPO method, and in Secs.~\ref{sec:spin_boson} and~\ref{sec:lz_prot} we benchmark the corrected PME expectation values against TEMPO 
for the spin-boson model and a LZ 
protocol, respectively. 
Finally, we briefly summarise our findings in Sec.~\ref{sec:conc}.

\section{Projection operators and the time-convolutionless master equation}\label{sec:TCL}

We begin with an overview of the Nakajima-Zwanzwig projection operator formalism~\cite{nakajima58,Zwanzwig60}
and the TCL master equation~\cite{breuer2002theory}, which will be useful for the discussions and derivations that follow in subsequent sections. 

We consider a system, $\mathcal{S}$,  
interacting with an environment, $\mathcal{E}$. 
The Hamiltonian is given by $H(t)= H_\mathrm{S}(t) + H_\mathrm{B} + H_\mathrm{I}$, where $H_\mathrm{S}(t)$ describes the system, which can in general be time dependent, $H_\mathrm{B}$ governs the free evolution of the environment, and  $H_\mathrm{I}$ describes the interaction between the two.  
We define an interaction picture with respect to $H_0(t)={H}_\mathrm{S}(t) + H_\mathrm{B}$. The evolution of the joint system-environment state $\chi(t)$ in this interaction picture is governed by the Liouville-von Neumman equation,
\begin{equation}
\partial_t\chi(t) =-i\left[H_\mathrm{I}(t), \chi(t)\right] \equiv\alpha\LL(t)\chi(t).
\end{equation}
The Liouvillian, $\mathcal{L}(t)$, is a superoperator that acts in the full Hilbert space of the system and environment, and thus leads to a computationally intractable problem in most practical circumstances (notice that we have introduced a parameterisation of the Liouvillian with $\alpha$, a dimensionless coupling parameter, which will be useful in the perturbative analysis later in this section). 
Instead, we wish to derive an equation of motion for the reduced state of the system $\rho_\mathrm{S}(t) = \tr_\mathrm{B}\left(\chi(t)\right)$.
To do this we define a projection operator $\PP$~\cite{nakajima58, Zwanzwig60}, whose action is $\PP\chi(t) = \rho_\mathrm{S}(t)\otimes\tau_\mathrm{B}$, where $\tau_\mathrm{B}$ is the reference state of the bath typically taken as the Gibbs state $\tau_\mathrm{B} = \exp(-\beta H_B)/\tr_\mathrm{B}(\exp(-\beta H_B))$ at inverse temperature $\beta=1/k_B T$. 
The projector is idempotent, $\PP^2 = \PP$, and we will refer to $\PP\chi(t)$ as the \emph{relevant} part of the density operator.
In addition, we also have an \emph{irrelevant} part of the density operator $\QQ\chi(t)$, where we have introduced the orthogonal projector $\QQ = \mathds{1}-\PP$.

With these projectors, we can define an equation of motion for both the relevant and irrelevant part of the evolution,
\begin{align}
\label{eq:relevant}
\partial_t\PP\chi(t) =& \alpha\PP\LL(t)\QQ\chi(t),\\
\label{eq:irrelevant}
\partial_t\QQ\chi(t) = &\alpha\QQ\LL(t)\PP\chi(t) + \alpha\QQ\LL(t)\QQ\chi(t),
\end{align}
where we have assumed that by construction $\PP\LL(t)\PP=0$~\cite{breuer2002theory}.
By integrating Eq.~\ref{eq:irrelevant}, we can find a formal solution for the irrelevant contribution of the form
\begin{equation}\label{eq:irrelevant_sol}
\begin{split}
\mathcal{Q} \chi(t)=&[1-\Sigma(t)]^{-1}\Sigma(t) \mathcal{P} \chi(t)\\
&+ [1-\Sigma(t)]^{-1} \mathcal{G}(t,t_0) \mathcal{Q} \chi(t_0), 
\end{split}
\end{equation}
where we have introduced the time-ordered Green's function for the irrelevant projection, 
\begin{equation}
\mathcal{G}(t,t_0) = \mathcal{T}_\leftarrow\exp\left(\alpha\int_{t_0}^t\QQ\LL(s) \operatorname{d}\!s\right).
\end{equation}
Here $\mathcal{T}_\leftarrow$ is the forward time-ordering operator, which orders operators with early times to the right.
We have also defined the superoperator, 
\begin{equation}
    \Sigma(t)=\alpha \int_{t_0}^t \dd{s} \mathcal{G}(t, s) \mathcal{Q} \mathcal{L}(s) \mathcal{P} {G}(t, s),\label{eq:sigma}
\end{equation}
which contains the reverse time-ordered Green's function,
\begin{equation}
G(t, s)=\mathcal{T}_{\rightarrow} \exp \left[-\alpha \int_s^t \dd{s}^{\prime} \mathcal{L}\left(s^{\prime}\right)\right],
\end{equation}
where the reverse time-ordering operator $\mathcal{T}_{\rightarrow}$ orders operators with early times on the left.

We can now expand Eq.~\ref{eq:irrelevant_sol} in terms of the dimensionless coupling parameter $\alpha$, which we truncate at second order to give
\begin{equation}\label{eq:irrel}\begin{split}
    \mathcal{Q}\chi(t)&=\QQ\chi(t_0) + \alpha \int^t_{t_0} \dd{s} \mathcal{L}(s)\mathcal{P}\chi(t) \\ &+\alpha^2\int^t_{t_0}\dd{s_1}\int^{s_1}_{t_0} \dd{s_2} \mathcal{Q}\mathcal{L}(s_1)\mathcal{L}(s_2)\mathcal{P}\chi(t)+\mathcal{O}(\alpha^3).
\end{split}\end{equation}
Substituting the above into Eq.~\ref{eq:relevant} and keeping terms only to second order in $\alpha$, we obtain the second-order TCL form
\begin{equation}\begin{split}
\partial_t\PP\chi(t) =& \alpha\PP\LL(t)\QQ\chi(t_0)\\ &+\alpha^2\int_{t_0}^t \operatorname{d}\!s\PP\LL(t)\LL(s)\PP\chi(t) + \mathcal{O}(\alpha^3).
\end{split}
\end{equation}
Finally, to get an equation of motion for the reduced system state we take a partial trace over the environmental degrees of freedom to obtain the second-order TCL master equation, written in the interaction picture as
\begin{equation}
\partial_t\rho_\mathrm{S}(t) = 
\mathcal{K}[\rho_\mathrm{S}(t)].
\end{equation}
Here we have introduced the dissipation superoperator
\begin{equation}\label{eq:NakaSchwan}
\mathcal{K}[\rho_\mathrm{S}(t)]=-\int_{t_0}^t\dd s \operatorname{tr}_\mathrm{B}\left[H_\mathrm{I}(t),\left[
H_\mathrm{I}(s),\rho_\mathrm{S}(t)\!\otimes\!\tau_\mathrm{B}
\right]\right],
\end{equation}
and we have chosen the bath initial state to match the reference state, such that $\mathcal{Q}\chi(t_0)=0$.

\section{The spin-boson model and Polaron theory}
\label{sec:polaron}
Having derived a general second-order TCL master equation, let us know consider a concrete physical model. 
We focus on the spin-boson Hamiltonian~\cite{breuer2002theory} 
describing a two-level system (TLS) interacting with a bosonic environment, 
given by ($\hbar=1$) 
\begin{equation}\label{eq:spin_boson}
\begin{split}
H(t) =& \frac{\epsilon(t)}{2}\hat\sigma_{z} + \frac{\Delta}{2}\hat\sigma_{x}
+ \hat\sigma_z\sum_k g_k (b_k^\dagger + b_k)
+\sum_k \nu_k b^\dagger_k b_k,
\end{split}\end{equation} 
where we have defined a static tunnelling parameter $\Delta$ and allowed 
the bias $\epsilon(t)$ to be potentially time-dependent. $\{\hat\sigma_j\}_{j=x,y,z}$ are the usual Pauli matrices.
The operator $b_k$ ($b_k^\dagger$) is the annihilation (creation) 
operator for the $k^{\mathrm{th}}$-mode of the environment, 
with $\nu_k$ the corresponding frequency and $g_k$ its coupling 
strength to the system. 
The overall coupling of the system to the environment is specified by the spectral 
density $J(\nu) = \sum_k \vert g_k\vert^2\delta(\nu-\nu_{k})$, 
which 
gives the interaction strength 
weighted by the environmental 
density of states. We shall refer to $H(t)$ as defining the Hamiltonian in the lab frame.

\subsection{Polaron transformation}

To account for strong system-environment coupling we apply the (active) polaron transformation to the lab-frame Hamiltonian $H(t)$ to give $H_\mathrm{P}(t)= e^{S}H(t)e^{-S}$, where 
$S = \hat\sigma_z\sum_k (g_k/\nu_k)(b_k^\dagger - b_k)$. This dresses the system states with vibrational modes of the environment, resulting in the polaron-frame Hamiltonian 
\begin{equation}\label{eq:polaron_ham}
\begin{split}
H_{\rm P}(t) =& \frac{\epsilon(t)}{2}\hat\sigma_{z} + \frac{\Delta_\mathrm{R}}{2}\hat\sigma_{x}
+\frac{\Delta}{2}(\hat\sigma_x\hat{B}_{x} + \hat\sigma_y \hat{B}_{y})\\
&+ \sum_k \nu_kb^\dagger_k b_k,
\end{split}
\end{equation}
where we have introduced the environment operators $\hat{B}_x = (\hat{B}_+ + \hat{B}_- - 2B)$/2 and $\hat{B}_y = i(\hat{B}_+ - \hat{B}_-)/2$, and defined the bath displacement operators,
\begin{equation}
\hat{B}_\pm = \exp(\pm\sum_k\frac{g_k}{\nu_k}(b^\dagger_k -b_k)).
\end{equation}
We have also defined the thermal expectation value of these
operators, given by
\begin{equation}\begin{split}
B &= \tr_B(\hat{B}_{\pm}\tau_\mathrm{B}),\\ 
&= 
\exp(-2\sum_k\frac{\vert g_k\vert^2}{\nu_k^2}\coth(\beta\nu_k/2)).
\end{split}
\end{equation}

There are several interesting features to the polaron transformed Hamiltonian. For example, the system-environment interaction term now induces transitions between the 
$\hat{\sigma}_z$ eigenstates of the TLS, with corresponding displacements of the phonon environment, in contrast to the linear coupling form of the original lab-frame Hamiltonian $H(t)$. 
In addition, the tunnelling is renormalised, $\Delta_{\rm R}= B \Delta$, where $B\leq1$ by definition. 
This is a consequence of bath interactions suppressing the tunnelling of the system due to the energetic costs of displacing the phonon environment as the system state varies.

\subsection{Polaron-frame master equation}
We can now derive a TCL master equation in the polaron frame using Eq.~\ref{eq:NakaSchwan} with 
the interaction Hamiltonian  
$H_\mathrm{I}(t) = \frac{\Delta}{2}(\hat\sigma_x(t)\hat{B}_x(t) 
+\hat\sigma_y(t)\hat{B}_y(t))$, written in the interaction picture with respect to the non-interacting terms in $H_{\rm P}(t)$.  
Note that as we have performed the polaron transformation, the projection operator 
$\mathcal{P}\chi(t)= 
\rho_\mathrm{S}(t)\otimes{\tau}_\mathrm{B}$ now assumes ${\tau}_\mathrm{B}$ to be the Gibbs state in the polaron frame. This will be important when considering observable expectation values below.  We assume that the phonon environment is initialised in the state ${\tau}_\mathrm{B}$
such that $\mathcal{Q}\chi(t_0) = 0$ and thus there are no 
inhomogeneous 
terms in the master equation. 

We find the polaron-frame dissipation superoperator to take the compact form
\begin{equation}\label{eq:pol_master}
    \mathcal{K}[\rho_\mathrm{S}(t)]=
    -\frac{\Delta^2}{4}\!\sum_{\alpha=x,y}\!
[\hat\sigma_\alpha(t),\hat\Theta_\alpha(t)\rho_\mathrm{S}(t)] 
+\mathrm{h.c.}
\end{equation}
where $\mathrm{h.c.}$ is the hermitian conjugate. 
Dissipation due to the phonon environment is induced through 
the non-Hermitian and non-Markovian rate operators, 
\begin{equation}
    \hat\Theta_\alpha(t)=\int_{t_0}^t
    \dd s\Lambda_{\alpha}(t-s)\hat\sigma_\alpha(s).
\end{equation}
Here we have introduced the polaron-frame environment 
two-time correlation functions, defined as
\begin{align}
    \Lambda_{x}(\tau) &=\langle \hat{B}_x(\tau)\hat{B}_x\rangle
    = \frac{B^2}{2}\left(e^{\phi(\tau)} + e^{-\phi(\tau)}-2\right),\\
    \Lambda_{y}(\tau)&= 
    \langle \hat{B}_y(\tau)\hat{B}_y\rangle
    =\frac{B^2}{2}\left(
e^{\phi(\tau)}
-e^{-\phi(\tau)}
    \right),
\end{align}
written in terms of the phonon propagator,
\begin{equation}
\phi(\tau) =\int\limits_0^\infty \frac{4J(\nu)}{\nu^2}\!\left(\!\coth(\frac{\beta\nu}{2}\!)\cos(\nu\tau) - i\sin(\nu\tau)\!\right)\dd\nu.
\end{equation}
In the following analysis, we shall assume the spectral density to take the form
\begin{equation}
J(\nu) = \alpha\omega_c^{-2}\nu^3 e^{-\nu/\omega_c},
\end{equation}
where $\alpha$ is a dimensionless coupling constant and $\omega_c$ is the cut-off frequency, which sets the response time of the phonon environment.

Eq.~\ref{eq:pol_master} provides a dynamical description of the reduced state of the system in the polaron frame, which treats the system-environment interaction 
non-perturbatively. 
Instead, the polaron-frame master equation can be thought of as perturbative in the ratio $\Delta/\omega_c$.
For more detailed discussion of the validity of the polaron approach, we refer the reader to Refs.~\cite{mccutcheon2010quantum} and \cite{Cosacchi21}.

 



\section{Extracting lab-frame observables using polaron theory}
\label{sec:lab_frame}
The master equation derived above gives the dynamics of the reduced system in the polaron frame. However, we are typically interested in system observables as defined in the physically accessible lab-frame. 
By construction, the polaron transformation is a unitary mapping that mixes system and environment degrees of freedom, meaning that observables as represented in the polaron frame do not directly correspond to those in the lab-frame, unless the observable commutes with the polaron transformation (for the spin-boson model, this is the case for example for $\hat\sigma_z$). 
Here we show that failing to account for the dressing induced by the polaron transformation when calculating lab-frame observables leads to deviations in the expected values of the system. 
We correct for these deviations by utilising the projection operator formalism introduced in Sec.~\ref{sec:TCL}.

We start by considering the relation between the lab and polaron frame density operators in the Schr\"odinger picture,
\begin{equation}
\chi(t) = e^{S} \chi_{\mathrm{L}}(t) e^{-S},
\end{equation}
where $\chi_{\rm L}(t)$ is the lab-frame density operator of the combined system and environment. 
Inverting this expression, and considering the expectation value of a system observable $A_{\rm L}$ in the lab frame, we have
\begin{equation}
\langle A_\mathrm{L}\rangle =
\tr_{\mathrm{S}+\mathrm{B}}(A_\mathrm{L}\chi_\mathrm{L}(t)) 
=\tr_{\mathrm{S}+\mathrm{B}}(A_\mathrm{L} e^{-S} \chi(t) e^{S}).  
\end{equation}
Using the cyclic invariance of the trace we obtain $\langle A_{\rm L}\rangle =\tr_{\mathrm{S}+\mathrm{B}}(A_{\rm P}\chi(t))$, 
where $A_{\rm P} = e^{S}A_{\rm L} e^{-S}$ is the polaron-frame operator, which  
can differ from $A_{\rm L}$ and potentially act on both the system and environment Hilbert space.

We only have access to the system Hilbert space through the TCL master equation. Nevertheless, to calculate lab-frame  expectation values we can make use of the projection operator by inserting the identity $\mathds{1} = \PP +\QQ$. This yields~\cite{kolli2011electronic} 
\begin{equation}
\langle A_\mathrm{L}(t)\rangle =\langle A_\mathrm{P}(t)\rangle_\mathrm{rel}
+\langle A_\mathrm{P}(t)\rangle_\mathrm{irrel},
\end{equation}
where $\langle A_\mathrm{P}(t)\rangle_\mathrm{rel}=\tr_{\mathrm{S}+\mathrm{B}}(A_{\rm P}\PP\chi(t))$ and $
\langle A_\mathrm{P}(t)\rangle_\mathrm{irrel}=\tr_{\mathrm{S}+\mathrm{B}}(A_{\rm P}\QQ\chi(t))$.
If $A_\mathrm{P}$ acts purely on the system Hilbert space, then by construction $\langle A_\mathrm{P}(t)\rangle_\mathrm{irrel}=0$. 
However, as we have stressed, 
system operators that do not commute with the polaron transformation are dressed by environmental displacement operators in the polaron frame.
In general, we can decompose such operators as
\begin{equation}
A_\mathrm{P} = \sum_\alpha \hat{s}_{\alpha}\otimes \hat{B}_\alpha,
\end{equation}
in terms of system operators $\hat{s}_\alpha$ and bath operators $\hat{B}_\alpha$. 
The relevant contribution to $\langle A_\mathrm{L}(t)\rangle$ is then 
\begin{equation}
    \langle A_\mathrm{P}(t)\rangle_\mathrm{rel} = \sum_\alpha
    \langle \hat{B}_\alpha\rangle
    \langle\hat{s}_\alpha(t)\rangle.
\end{equation}
This is the form of expectation value defined in previous studies of the PME~\cite{mccutcheon2010quantum,nazir2016modelling}.

For the irrelevant contribution to $\langle A_\mathrm{L}(t)\rangle$, we can make use of the expansion of the irrelevant part of the density operator given Eq.~\ref{eq:irrelevant_sol} to find
\begin{equation}\begin{split}
\langle A_\mathrm{P}(t)\rangle_\mathrm{irrel}&\approx\alpha \int^t_{t_0} \dd{s}
\tr_\mathrm{S+E}(A_\mathrm{P}(t)\mathcal{L}(s)\mathcal{P}\chi(t))\\
&=-i \int^t_{t_0} \dd{s}\tr_\mathrm{S+E}\left(
A_\mathrm{P}(t)
[H_\mathrm{I}(s), \rho_\mathrm{S}(t)\otimes\tilde{\tau}_\mathrm{B}]\right),
\label{eq:irr_expec}
\end{split}\end{equation}
where we have kept only the first-order correction. 
While this term originates from what is often referred to as the \emph{irrelevant} contribution in the projection operator formalism, we shall see that can have a significant impact on system expectation values.

The resulting correction terms are identically zero in the case that the system observables commute with the polaron transformation. 
Therefore, the population difference, $\langle\hat{\sigma}_z\rangle$, is the same in both the lab and polaron frames.
In contrast, operators involving the coherences are not. For example, $\hat{\sigma}_\mathrm{P} = \hat{\sigma}\otimes \hat{B}_+$ when $\hat{\sigma}_{\rm L}=\hat{\sigma} = \dyad{e}{g}$.
In this case, we have $\langle\hat{\sigma}_\mathrm{P}(t)\rangle_\mathrm{rel} =B\langle\hat{\sigma}(t)\rangle_\mathrm{P}$, where $\langle\cdot\rangle_\mathrm{P}$ denotes an expectation value taken in the polaron frame,
and the correction term is given by 
\begin{equation}\label{eq:irel_op_form}\begin{split}
    \langle\hat{\sigma}_\mathrm{P}(t)\rangle_\mathrm{irrel} 
&=
\tr_\mathrm{S}\left[\hat{\sigma}(t)\left(
\hat{\Psi}(t)\rho_\mathrm{S}(t) - \rho_\mathrm{S}(t)\hat{\Phi}(t)
\right)
\right],
\end{split}
    \end{equation}
where we have introduced the operators $\hat{\Psi}(t) = -i\Delta(\hat\Theta_x(t) + i\hat\Theta_y(t))/2$
and $\hat\Phi(t) = -i\Delta(\hat\Theta^\dagger_x(t)+i\hat\Theta_y^\dagger(t))/2$.

\section{Time Evolved Matrix Product Operator benchmarks}\label{sec:tempo}

We shall benchmark the corrections derived above against 
TEMPO~\cite{strathearn2018efficient}, which is naturally formulated within the lab frame. 
\subsection{The TEMPO formalism }

TEMPO is a reformulation of the discrete-time path integral approach originally developed by Makri and Makarov~\cite{makri1995tensorI,makri1995tensorII} (known as the Quasi Adiabatic Path Integral (QuAPI) technique) as a tensor network. 
To derive the TEMPO method, our starting point is the reduced density matrix of the system after $N$ time-steps $t = N\delta$, where $\delta$ is the step-size.
We use Liouville-space notation, defining a compound index $s = (s_l, s_r)$, which labels the left and right eigenstates of the system-environment coupling operator, which for the spin-boson model in Eq.~\ref{eq:spin_boson} is $\hat{\sigma}_z$.
In this notation, the components of the reduced density operator can be written as $\rho^{s} = \langle s_l\vert\rho\vert s_r\rangle$.
For a Gaussian environment, the dynamics of the reduced state can be expressed in terms of the discrete-time path integral~\cite{makri1995tensorI, makri1995tensorII, cygorek2022simulation},
\begin{equation}
\rho^{s_N}(N\delta) = \sum_{s_1\cdots s_N}\mathcal{F}^{s_1\cdots s_{N-1}}\left(\prod\limits_{l=1}^n \mathcal{M}_{s_l, s_{l-1}}\right)\rho^{s_0}(0),
\end{equation}
where $\mathcal{M}_{s_l, s_{l-1}}= (e^{\mathcal{L}_{l\delta}\delta})_{s_l, s_{l-1}}$ describes the free evolution of the system under the Liouvillian $\mathcal{L}_{t_i}[\rho] = -i[H_S(t_i), \rho]$.

The tensor $\mathcal{F}^{s_1\cdots s_{N-1}}$ is the discrete-time Feynman-Vernon influence functional~\cite{feynman63functional}, a rank-$N$ tensor that captures exactly the environmental influence in the limit that $\delta\rightarrow0$.
For the spin-boson model in which the environment is initialised in a thermal state, the influence functional can be decomposed as~\cite{makri1995tensorI} 
\begin{equation}~\label{eq:IF}
\mathcal{F}^{s_N\cdots s_1} = \prod\limits_{i=1}^n
\prod\limits_{j=1}^i [b_{i-j}]^{s_i,s_j},
\end{equation}
where we have introduced the tensors $[b_{i-j}]^{s_i, s_j}$, which may be written in terms of the bath correlation functions as
\begin{equation}
[b_{i-j}]^{s, s^\prime} = e^{-(\lambda_{s_l} - \lambda_{s_r})(\eta_{i-j}\lambda_{s^\prime_l} 
-\eta_{i-j}^\ast\lambda_{s^\prime_r})}.
\end{equation}
Here $\lambda_s$ are the eigenvalues of the system-environment coupling operator, and the memory kernel is defined as 
\begin{equation}
    \eta_{i-j} = \left\{
    \begin{array}{cc}
    \int_{t_{i-1}}^{t_i}dt^\prime\int_{t_{i-1}}^{t_i}dt^{\prime\prime} C(t^\prime- t^{\prime\prime}), & i=j,\\
    \int_{t_{i-1}}^{t_i}dt^\prime\int_{t_{j-1}}^{t_j}dt^{\prime\prime} C(t^\prime- t^{\prime\prime}),
    &
    i\neq j,
    \end{array}
    \right.
\end{equation}
where $C(t) = \int_0^\infty d\omega J(\omega)(\coth(\beta\omega/2)\cos(\omega t) - i\sin(\omega t))$ is the bath correlation function.

Due to its high-rank form, it is not feasible to calculate the influence functional directly, limiting to studies with few timesteps or for environments with short memories. 
However, Strathearn~\emph{et al.}~\cite{strathearn2018efficient} noticed that Eq.~\ref{eq:IF} could be recast as a tensor network and propagated using matrix product operator methods. Notably, this allows one to use standard tensor compression methods based on singular value decomposition, which reduces the rank of the influence functional. 
Since this initial proposal there has been a series of contraction and compression schemes~\cite{jorgensen2019exploiting,cygorek2024sublinear,link2023open} that improve upon the original formulation.
The benchmarks in this work utilise the scheme developed by Link~\emph{et al.}~\cite{link2023open} which make use of techniques from infinite uniform matrix product states~\cite{Vidal07itebd} to efficiently compress and contract the tensor network. For details of the scheme, we refer the reader to Ref.~\cite{link2023open}.

\begin{figure}[b]
    \centering
    \includegraphics[width=\columnwidth]{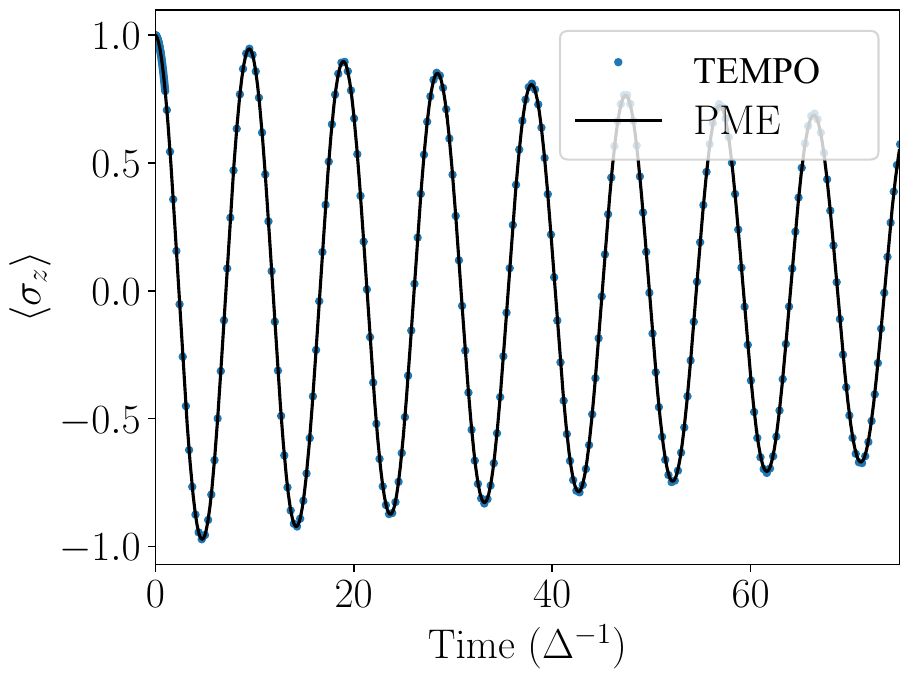}
    \caption{Dynamics of the population difference in the spin-boson model calculated with the PME (solid curve) and TEMPO (points). Parameters used in this plot are $\epsilon=0$, $\beta\Delta=1$, $\alpha = 0.2\Delta$, and $\omega_c=10\Delta$.}
    \label{fig:sz_expec}
\end{figure}

\subsection{Simulation details and sources of error}
The above formulation of TEMPO takes the environment to be initialised in 
thermal equilibrium in the lab frame~\cite{Otterpohl22hidden}. 
However, this differs from the initial state of the PME, which assumes a Gibbs state in the polaron
frame.
We can ensure the initial bath state to be the same in both theories 
in two ways. The first would be to assume that the initial state in the 
PME is a Gibbs state in the lab frame, this would 
introduce inhomogeneous terms to the master equation since then 
$\mathcal{Q}\chi(t_0)\neq0$.
Alternatively, we can allow TEMPO to thermalise to a Gibbs state 
in the polaron frame by initialising the system in an eigenstate of  
$\hat\sigma_z$ with no tunneling ($\Delta = 0$) 
for a time window $t_\mathrm{th}$, which exceeds the correlation time 
of the environment.
Since inhomogeneous terms may mask the effect of the correction 
contributions to the dynamics of system observables, we opt for the latter, 
where we set the thermalisation time in TEMPO to be $t_\mathrm{th} = 4\Delta^{-1}$ (using the non-zero value of $\Delta$ taken for the subsequent dynamics). This is also the more physically relevant initial condition if we assume that the system-environment interaction is not controllable, i.e.~it cannot be switched on or off.

\begin{figure*}[t]
    \centering
    \includegraphics[width=\textwidth]{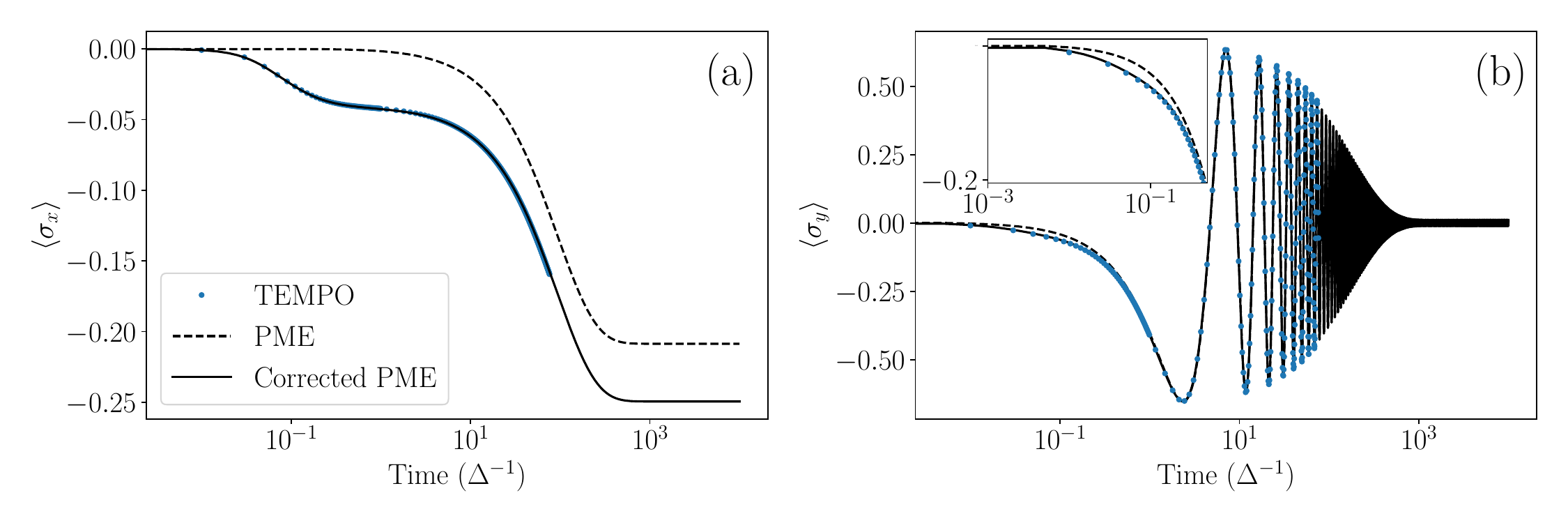}
    \caption{A comparison of the coherence dynamics calculated using the PME with (solid curves) and without (dashed curves) the expectation value correction terms 
    of Eq.~\ref{eq:SB_irrel2}. The dynamics have been benchmarked against TEMPO (points), where we have presented TEMPO simulations only up to $t=100$, after which the accumulation of Trotter errors makes convergence challenging. The inset in (b) shows a magnification at short times. Parameters used in these plots are $\epsilon=0$, $\beta\Delta=1$, $\alpha = 0.2\Delta$ and $\omega_c=10\Delta$.}
    \label{fig:coherence_dynamics}
\end{figure*}

When benchmarking the polaron method 
it is crucial to consider potential sources of error in the TEMPO simulations. 
The two main contributions are the Trotter error and the threshold 
for discarding singular values during tensor network compression. 
The Trotter error emerges due to the discretisation of the time axis 
using a second-order Trotter splitting, leading to an error of 
$\mathcal{O}(\delta^3)$. Consequently, $\delta$ must be chosen to be 
small relative to the system's energy scales, though even then Trotter error can accumulate over time (see below). 
The threshold $\varepsilon$ at which 
singular values are discarded 
is tuned to ensure the resulting system dynamics are numerically 
converged, typically set at $\varepsilon=10^{-9}$ for the models 
considered in this work.

\section{Static spin-boson model}
\label{sec:spin_boson}

We begin our analysis of the system dynamics by considering a spin-boson model with constant system parameters (set after $t_{\rm th}$ in TEMPO), where $\epsilon(t) = \epsilon\geq0$ and $\Delta>0$. 
In this case, the polaron-frame system Hamiltonian has eigenstates $\ket{\psi_{\pm}} = ((\epsilon \pm \eta)\ket{g} + \Delta\ket{e})/\sqrt{2\eta(\eta \pm \epsilon)}$, where $\eta^2 =\epsilon^2 + \Delta_\mathrm{R}^2$, which satisfy 
$H_\mathrm{S}\ket{\psi_\pm}=\pm(\eta/2)\ket{\psi_\pm}$.

We are interested 
in the impact that the correction (irrelevant) terms have on the expectation values of system observables over time.
Since the polaron transformation commutes with the system $z$-basis populations, the population difference is invariant between frames, that is $\langle\sigma_{z\mathrm{L}}(t)\rangle=\langle\sigma_{z\mathrm{P}}(t)\rangle$. 
Fig.~\ref{fig:sz_expec} shows an example of the population difference as a function time predicted by the PME (solid curve) and TEMPO (points). 
Here we see excellent agreement between the two methods, 
demonstrating that the PME accurately captures the system population dynamics in this regime.

If we consider the coherence however, as discussed earlier the lab-frame operator $\sigma_\mathrm{L} =\sigma= \ket{g}\!\bra{e}$ maps to the polaron-frame operator $\sigma_\mathrm{P} = \sigma\otimes B_+$. 
The presence of the bath operator $B_+$ necessitates the inclusion of a correction term,
such that
\begin{equation}\label{eq:SB_irrel}
    \langle\sigma_\mathrm{L}(t)\rangle 
    =
    B\langle \sigma(t)\rangle + \langle\sigma_\mathrm{P}(t)\rangle_\mathrm{irrel}.
\end{equation}
We can use the expression given in Eq.~\ref{eq:irel_op_form} to write the correction term as 
\begin{equation}\label{eq:SB_irrel2}
    \langle\sigma_\mathrm{P}(t)\rangle_\mathrm{irrel}=\langle\sigma(t)\hat\Psi(t)\rangle -\langle\hat\Phi(t)\sigma(t)\rangle.
\end{equation}
The Pauli operators can then be constructed from $\langle\sigma_x\rangle = 2\operatorname{Re}[\langle\sigma_\mathrm{L}(t)\rangle]$ and $\langle\sigma_y\rangle = -2\operatorname{Im}[\langle\sigma_\mathrm{L}(t)\rangle]$.

The comparison of TEMPO and PME-predicted coherence dynamics with and without correction terms is shown in Fig.~\ref{fig:coherence_dynamics}. Focusing on $\langle\sigma_x\rangle$, we see that TEMPO predicts a notable initial change in coherence (also observable in $\langle\sigma_y\rangle$). Without correction terms, the PME misses this short-time generation of coherence entirely, resulting in significant discrepancies with TEMPO and the corrected PME predictions for both the short- and long-time behaviour. 
Importantly, with correction terms, the PME closely reproduces the TEMPO dynamics. 
Notably, TEMPO simulations in this strong-coupling regime are computationally demanding, 
which here limits the propagation time 
after which the accumulation of Trotter errors makes the TEMPO results untrustworthy. 
Conversely, the PME is a computationally efficient method, 
facilitating the simulation of both short- and long-time dynamics, even with multiple timescales present. 

We attribute the initial generation of coherence in TEMPO and the corrected PME predictions to the dynamic dressing of the TLS states by phonon mode displacements. The \emph{relevant} contribution to the system expectation values captures the static dressing of the TLS, where it describes the stationary polarons associated with the ground and excited state configurations of the TLS. 
The \emph{irrelevant} (correction) contribution captures the effect of transitions between these polaron states due to tunneling, where the dressing of the system states changes dynamically with the state of the system itself. 
This provides an additional contribution to the coherence which accumulates on the timescale of polaron formation.

Similar phenomena occur in the context of the optical properties of solid-state quantum emitters. Here one observes rapid relaxation of the two-time optical coherence on picosecond timescales~\cite{brash2019light,brash2023nanocavity} as the optical fields induce transitions between the electronic configurations of the emitter, and thus the corresponding polaron states.
The resulting dynamics leads to the emergence of phonon sidebands in the emission spectrum~\cite{iles2017limits, iles2017phonon}. While these are different dynamical quantities, the underlying physical processes are the same.
More recently, Wiercinski \emph{et al}. \cite{wiercinski2024role} highlighted a discrepancy between coherence dynamics of superradiant emitters calculated through a polaron master equation and TEMPO, where TEMPO predicted a change in the emitter coherence on a picosecond timescale. 
Wiercinski \emph{et al}. accounted for this coherence slip by inverting the polaron transformation,  
which was possible for the considered system as there was no coherent driving present. 
In our case, the projection operator approach we have employed provides a systematic method to capture dynamic correction terms independent of the system or driving protocol considered, including 
the study of time-dependent drivings as will be outlined in the next section.

\begin{figure}[t!]
    \includegraphics[width=\columnwidth]{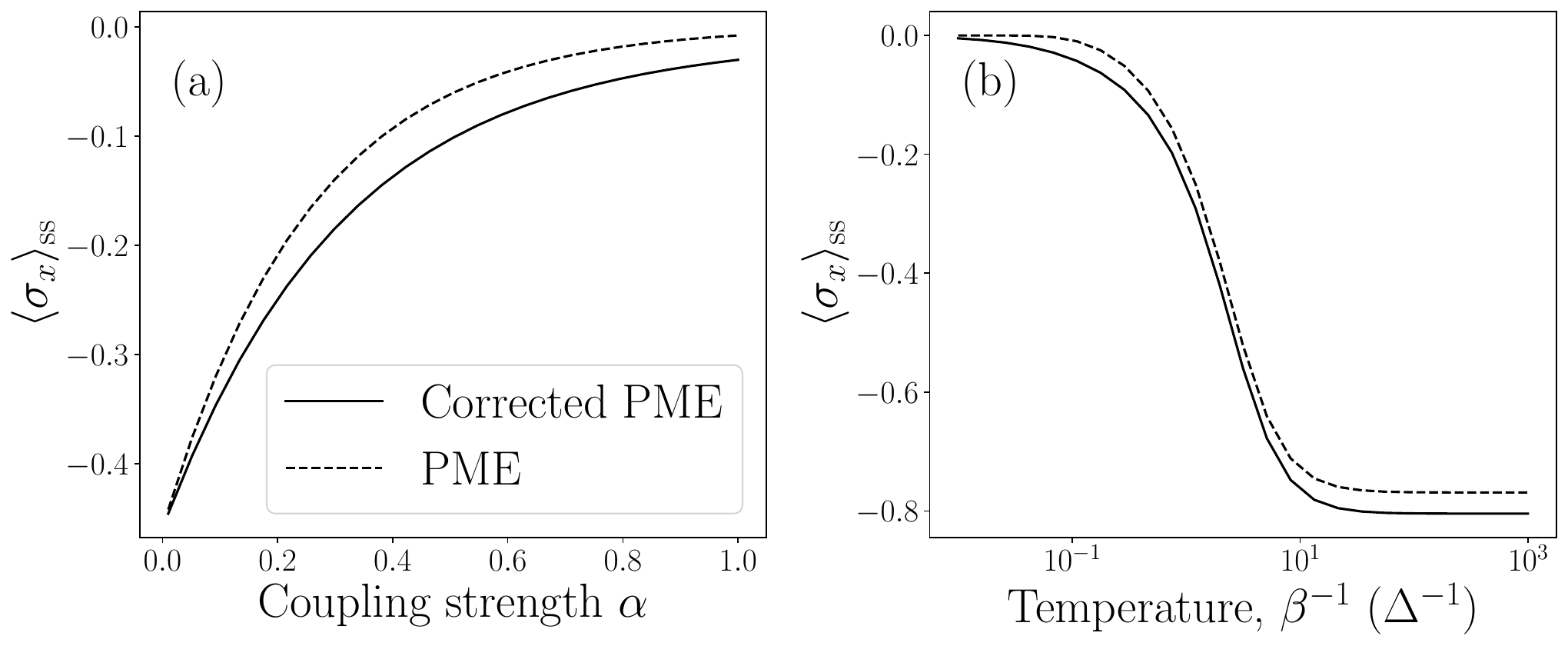}
    \caption{Steady-state coherence of the spin-boson model calculated using the PME with (solid curves) and without (dashed curves) correction terms for (a) increasing coupling strength, and (b) varying temperature. 
    Note that the system parameters match those used in Fig.~\ref{fig:coherence_dynamics}, while we have used $\beta\Delta=1$ for (a) and $\alpha=0.2\Delta$ for (b).}
    \label{fig:ss_coh}
\end{figure}

We can also consider the impact that the correction terms have on steady-state expectation values. Fig.~\ref{fig:ss_coh} shows the steady-state coherence calculated from the PME with and without correction terms for different system-environment coupling strengths and temperatures.
Here we see that the PME without correction terms underestimates the magnitude of the steady-state coherence at all non-zero values of the system-environment coupling strength. These differences become significant at intermediate and strong coupling, across a range of temperatures.

\section{Landau Zener transition}
\label{sec:lz_prot}

\begin{figure}[t]
    \centering
    \includegraphics[width=\columnwidth]{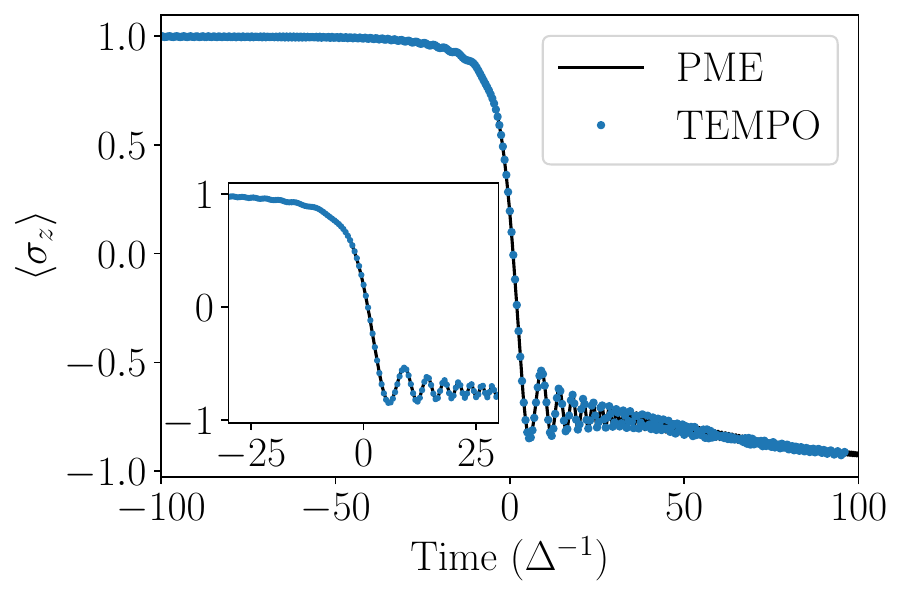}
    \caption{Population dynamics during a LZ protocol, calculated from the PME (solid curves) and TEMPO (points). There is a sign change in the bias as the system crosses the $t=0$ point, where an avoided crossing occurs in the system energy levels. The inset shows a magnification around the avoided crossing point. The parameters used here are $\nu = 0.1\Delta$, $\beta\Delta=1$, $\alpha = 0.4$, and $\omega_c = 10\Delta$.}
    \label{fig:lz_sz}
\end{figure}

\begin{figure*}
    \centering
    \includegraphics[width=\textwidth]{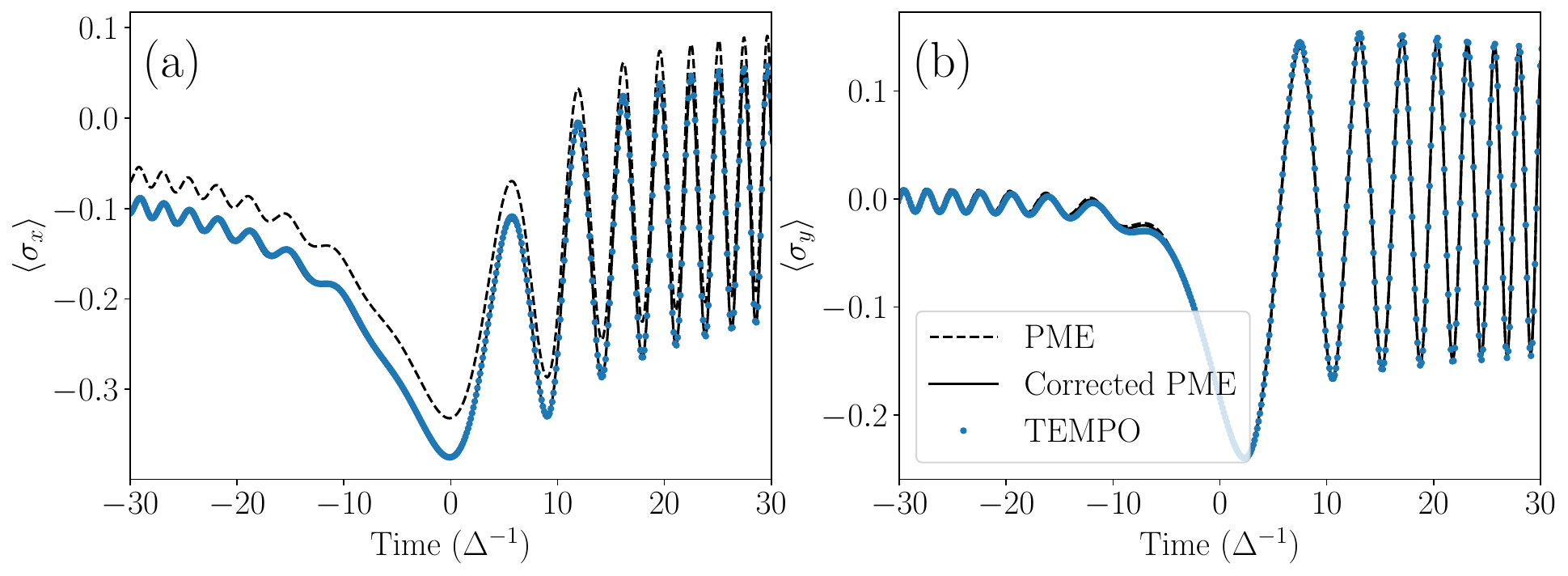}
    \caption{LZ coherence dynamics focused around the avoided crossing at $t=0$, calculated using TEMPO and the PME with (solid curves) and without (dashed curves) expectation value correction terms. 
    Notably, there exists a consistent offset in $\langle\sigma_x\rangle$ in the uncorrected PME, which is effectively addressed by including the appropriate corrections. The parameters used here are the same as in Fig.~\ref{fig:lz_sz}.}
    \label{fig:lz_coherence}
\end{figure*}

As a further application we now consider the effect of the polaron correction terms for the paradigmatic dissipative LZ protocol~\cite{wubs06lz,Saito07lz}.
In this setting the TLS has a fixed tunneling $\Delta$, but a bias that changes linearly with time according to $\epsilon(t) = \nu t$, where $\nu$ gives the rate at which the bias is varied and $t\in[t_i,t_f]$. Here we have introduced the initial and final protocol times $t_i$ and $t_f$, respectively, set such that $\epsilon(t)$ dominates over $\Delta$ at the begining and end of the protocol. 
At time $t=0$ there is no bias and the system eigenstates are governed by the tunneling $\Delta$, which yields an avoided crossing.

As is typical in a LZ protocol we take the sweep rate $\nu$ to be slow compared to the tunnelling rate $\Delta$. 
In the absence of coupling to a dissipative environment, this condition allows population to be (almost) adiabatically transferred between the ground state at $t_i$ (large negative bias, implying $\langle\sigma_z\rangle\approx1$) and the ground state at $t_f$ (large positive bias, implying $\langle\sigma_z\rangle\approx-1$).
However, in the presence of a dissipative environment, phonon induced transitions can drive the system population between the ground and excited state even in the strictly adiabatic limit.
This leads to imperfect population transfer at the end of the protocol~\cite{wubs06lz,diba2024quantum}.

The assumption of an adiabatic LZ protocol also allows us to simplify the form of 
the PME 
by making an additional Markov approximation.
This is done by taking the initial time $t_0\rightarrow-\infty$ in the integrals of Eq.~\ref{eq:pol_master}.
This simplifies the interaction picture rate operators in Eq.~\ref{eq:pol_master} to
\begin{equation}
\hat{\Theta}_\alpha(t)\approx\int_0^{\infty}\dd\tau~\Lambda_\alpha(\tau)\hat{\sigma}_\alpha(t-\tau),
\end{equation}
where we have used the substitution $s=t-\tau$. Note that a Markov approximation within the polaron frame still naturally accounts for some non-Markovian features observed within the lab frame~\cite{mccutcheon2010quantum}, and thus is not as restrictive as the lab-frame Markov approximation made in a weak-coupling Redfield theory~\cite{Redfield65}.

Fig.~\ref{fig:lz_sz} demonstrates this, showing excellent agreement over the full protocol for the population dynamics  
calculated through both the PME and TEMPO in the strong system-environment coupling regime. 
The two level system is initially prepared in the positive eigenstate of the $\hat\sigma_z$ operator, which is very close to the system ground state at large negative bias. The bias is then linearly swept and, as expected,  
the presence of the strongly-coupled environment leads to imperfect population transfer as the system traverses the avoided crossing,  
with the final system population difference 
$\langle\hat\sigma_z\rangle\neq-1$.

The system coherence is illustrated in Fig.~\ref{fig:lz_coherence}. 
As in the spin-boson model, there is a notable difference in the expectation value 
$\langle\sigma_x\rangle$ 
between the PME without corrections and TEMPO, which persists throughout the protocol. 
This again arises from the short-time dynamical dressing of the TLS states due to the presence of the tunneling parameter $\Delta$, 
which is fully accounted for by the dynamical correction terms, leading to excellent agreement with TEMPO. 
Notably, at the beginning of the protocol, the system's behavior is dominated by the bias, as 
 $\vert\epsilon(t_i)\vert\gg\Delta$. 
 However, even in this case, dynamical dressing introduces observable differences in coherence throughout the protocol, demonstrating the robustness of the effect across varying parameter regimes.

\section{Conclusion}\label{sec:conc}
In conclusion, our study revisits the PME and extends its application to modeling the non-Markovian dynamics of open quantum systems strongly coupled to their environments. Although the PME effectively predicts population dynamics, its inability to fully capture coherences limits its overall accuracy. This issue arises because of the mixing of system and environment degrees of freedom introduced by the transformation to the polaron frame, which we have shown impacts the precision of calculated expectation values. By employing the Nakajima-Zwanzig projection operator formalism, we introduced correction terms that significantly enhance the description of system coherences (or more generally, system observables that do not commute with the polaron transformation).

We demonstrated the importance and effectiveness of these correction terms through two specific examples, the canonical spin-boson model and a dissipative time-dependent Landau-Zener protocol. In both cases, the corrected PME showed an improved (almost perfect) agreement with numerically exact methods, validating our approach.
Our findings highlight the need for incorporating such corrections in PME applications to achieve more accurate and reliable descriptions of open quantum system dynamics. For example, future work could extend the correction terms to study the impact of dynamical phonon dressing on multi-time correlation functions,  
which play a key role in determining the spectral 
and coherence properties of solid-state quantum emitters.

\bibliography{apssamp}
\end{document}